\documentclass{ifacconf}
\usepackage{natbib}        
\usepackage{stfloats}
\usepackage{graphicx}      
\usepackage{booktabs}
\usepackage{amsmath,amssymb,amsfonts}
\usepackage{verbatim}
\usepackage{dsfont}

\newtheorem{thm}{Theorem}
\newtheorem{lem}{Lemma}
\newtheorem{defn}{Definition}
\newtheorem{exmp}{Example}
\newtheorem{rem}{Remark}
\newtheorem{assum}{Assumption}
\newtheorem{prob}{Problem}
\newtheorem{prop}{Proposition}
\newtheorem{cor}{Corollary}
\newtheorem{fact}{Fact}
\begin{document}
\begin{frontmatter}

\title{A Scenario Approach to the Robustness of \mbox{Nonconvex--Nonconcave}~Minimax~Problems\thanksref{footnoteinfo}} 

\thanks[footnoteinfo]{
This work was supported by the Swedish Research Council Distinguished
Professor Grant 2017-01078, the Knut and Alice Wallenberg Foundation
Wallenberg Scholar Grant, and the Swedish Strategic Research Foundation
SUCCESS Grant FUS21-0026.
The authors are also affiliated to Digital Futures. \textit{Corresponding author: G. Chen.}}

\author[KTH]{Huan Peng} 
\author[SEU]{Guanpu Chen} 
\author[KTH]{Karl H. Johansson} 

\address[KTH]{School of Electrical Engineering and Computer Science, \\
KTH Royal Institute of Technology, SE-100 44, Stockholm, Sweden \\
e-mail: \{huanp, kallej\}@kth.se } 
\address[SEU]{School of Automation, Southeast University, Nanjing 210096, China \\
e-mail: guanpu\_chen@seu.edu.cn }

\begin{abstract}                
This paper investigates probabilistic robustness of nonconvex--nonconcave minimax problems via the scenario approach. Specifically, under convex strategy sets for all players, inspired by recent advances in scenario optimization, we first establish a probabilistic robustness guarantee for an $\varepsilon$-stationary point, overcoming the dependence on the non-degeneracy assumption by proving the monotonicity of the stationary residual in the number of scenarios. Furthermore, in the presence of nonconvex strategy sets, we reveal the fundamental difficulty of obtaining a tight theoretical bound based on this recent framework. Consequently, we establish a relaxed, yet rigorously valid, probabilistic bound for a global minimax point. A numerical experiment corroborates our theoretical findings.
\end{abstract}


\end{frontmatter}

\section{Introduction}
Minimax problems model a two-player zero-sum game where one player aims to minimize a payoff function while the other aims to maximize it. This structure underpins a vast array of important applications, from optimal control~\citep{bacsar2008h} to machine learning~\citep{Sutton2018}. For the convex case, the minimax problem is well-studied~\citep{v1928theorie, fan1953minimax, sion1958general}. Recent advances in machine learning have spurred growing interest in nonconvex minimax problems. Prominent examples include generative adversarial networks (GANs)~\citep{goodfellow2014generative} and adversarial training~\citep{madry2018towards}, which can often be formulated as nonconvex minimax problems. 

Uncertainty affects the decision-making process in many applications. The scenario approach provides a data-driven methodology for addressing uncertainty: it yields probabilistic robustness guarantees without requiring prior knowledge of the probability distribution or geometric structure of the uncertainty set. Building on this approach, a substantial body of work investigated uncertain convex games, e.g., \citet{fele2020probably,chen2025inverse}. Most existing analyses, however, critically rely on the assumption of convexity; when this assumption is violated, the resulting bounds may not be directly applied. To address this limitation, \cite{campi2018general} first studied general nonconvex scenario optimization, while \cite{garatti2025non} subsequently achieved significantly tighter risk bounds across a broad class of problems, albeit with more technically involved derivations.

The probabilistic robustness of minimax problems has typically been studied under restricted settings. Existing studies, whether convex \citep{doi:10.1137/130928546} or nonconvex \citep{assif2020scenario,garatti2025non}, are limited to the case where the cost function is subject to uncertainty. This setting is typically dealt with by worst-case reformulation, which is equivalent to a minimax problem. A critical gap remains, namely, when the strategy sets of both players are affected by uncertainty. Addressing this problem, particularly when the payoff is nonconvex--nonconcave with respect to the two players’ strategies, is mathematically challenging. Moreover, a Nash equilibrium (NE) in this case may fail to exist or may lack standard well-posedness properties. Therefore, the robustness of equilibria under these conditions remains an open question.

In this paper, we consider nonconvex--nonconcave minimax problems in which both players’ strategy sets are subject to uncertainty. The primary challenge lies in establishing probabilistic robustness guarantees without the non-degeneracy assumption. We address this challenge by characterizing the robustness of key equilibria through two distinct pathways. We leverage the latest developments in scenario optimization~\citep{garatti2025non} to handle cases with convex strategy sets, while utilizing an alternative theoretical framework~\citep{campi2018general} to secure robustness guarantees in fully nonconvex settings.

The main contributions of this paper are twofold. Firstly, we establish the robustness of an $\varepsilon$-stationary point, the first-order condition for a local NE, under nonconvex--nonconcave payoffs and convex strategy sets. By demonstrating that the stationary residual is non-increasing as the number of scenarios increases, we derive a tight probabilistic robustness guarantee for the $\varepsilon$-stationary point~(Theorem~\ref{thm:epsilon}). Secondly, to further investigate the case with nonconvex strategy sets, we derive a probabilistic robustness guarantee for a global minimax point by explicitly accounting for the sequential order of the players' decision-making. Leveraging the extreme value theorem, we point out the fundamental difficulty in establishing a tight bound, and consequently, provide an alternative, formally valid robustness bound for this fully nonconvex setting~(Theorem~\ref{thm:nonconvex}).

The remainder of the paper is organized as follows. Section~\ref{sec:The Scenario Minimax Problem} formulates the minimax problem and revisits a robustness result for NE in the convex--concave case. Section~\ref{sec:nonconvex} establishes probabilistic robustness guarantees for an $\varepsilon$-stationary point and a global minimax point. Section~\ref{sec:Numerical Example} presents a numerical experiment, while Section~\ref{sec:conclusion} concludes the paper.

\section{The Scenario Minimax Problem}   \label{sec:The Scenario Minimax Problem}
In this section, we introduce the minimax problem and revisit a robustness result from the literature.

\subsection{Problem Formulation}
Let $\theta$ be a random parameter taking values in a set $\Theta$, equipped with a $\sigma$-algebra $\mathcal{F}$, and let $\mathbb{P}$ be the associated probability measure on $\mathcal{F}$. Let $\mathcal{X}_{\theta} \subseteq \mathbb{R}^{p}$ and $\mathcal{Y}_{\theta} \subseteq \mathbb{R}^{q}$ be $\theta$-dependent uncertain strategy sets, and let $f:\mathcal{X}_{\theta} \times \mathcal{Y}_{\theta} \to \mathbb{R}$ be a continuously differentiable payoff function. The minimax problem
\begin{equation}   \label{eq:minimax_def}
    \min_{x \in \mathcal{X}_{\theta}} \max_{y \in \mathcal{Y}_{\theta}} f(x,y)
\end{equation}
represents a two-player zero-sum game under uncertainty. The min-player selects a strategy $x \in \mathcal{X}_{\theta}$ to minimize the payoff, whereas the max-player selects a strategy $y \in \mathcal{Y}_{\theta}$ to maximize it.

Generally, one cannot expect a solution to \eqref{eq:minimax_def} unless the set $\Theta$ is exactly known or additional assumptions on the distribution $\mathbb{P}$ are imposed, e.g.,~\citet{chen2021distributed,fochesato2023generalized}. To overcome this obstacle, we employ a data‑driven framework, replacing \eqref{eq:minimax_def} with its empirical approximation based on
an independent and identically distributed (i.i.d.) multi-sample consisting of $M$ samples (also called ``scenarios'') $\boldsymbol{\theta}^{M} = (\theta_{1},\dots,\theta_{M}) \in \Theta^{M}$ drawn from $\Theta$. Since the scenarios are drawn independently, $\boldsymbol{\theta}^{M}$ is defined on the product probability space $(\Theta^{M},\mathcal{F}^{M},\mathbb{P}^{M})$. This leads to the following deterministic \emph{scenario minimax problem}:
\begin{equation}   \label{eq:scenario_minimax_def}
    \min_{x \in  \mathcal{X}^M}
    \max_{y \in \mathcal{Y}^M}
    f(x,y),
\end{equation}
where $\mathcal{X}^M := \bigcap_{i = 1}^{M} \mathcal{X}_{\theta_{i}}$ and $\mathcal{Y}^M := \bigcap_{i = 1}^{M} \mathcal{Y}_{\theta_{i}}$, provided that $\mathcal{X}^M \times \mathcal{Y}^M$ is nonempty for any $M \in \mathbb{N}$.

We first revisit the convex--concave case, where $f$ is convex in $x$ and concave in $y$. In this setting, the scenario minimax problem \eqref{eq:scenario_minimax_def} is well-studied~\citep{v1928theorie, fan1953minimax, sion1958general}, with NE serving as a central concept.

\begin{defn}
    A collective strategy $(x^{*},y^{*})$ is an NE of problem \eqref{eq:scenario_minimax_def} if, for all $(x,y) \in \mathcal{X}^M \times \mathcal{Y}^M$:
    \begin{equation}   \nonumber
        f(x^*,y) \le f(x^*,y^*) \le f(x,y^*).
    \end{equation}
\end{defn}
In the convex--concave case, the NE is equivalent to a global saddle point, satisfying the condition that no player has incentive to unilaterally deviate. The following assumption is widely employed to address uncertainty in strategy sets~\citep{chen2021distributed,fochesato2023generalized,xu2023algorithm}.
\begin{assum}  [Strategy convexity]   \label{asp:nonempty}
    $\mathcal{X}_{\theta}$ and $\mathcal{Y}_{\theta}$ are convex, compact, and nonempty for all $\theta \in \Theta$.
\end{assum}

\begin{lem} [{\citealp{fan1953minimax}}] \label{lemma:convex_existence}
    Supposing $f$ is convex--concave. Under Assumption \ref{asp:nonempty}, the scenario minimax problem \eqref{eq:scenario_minimax_def} admits an NE on $\mathcal{X}^M \times \mathcal{Y}^M$. Furthermore, the minimax equality holds:
    \begin{equation}  \nonumber
        \min_{x \in \mathcal{X}^M} \max_{y \in \mathcal{Y}^M} f(x,y) 
        = \max_{y \in \mathcal{Y}^M} \min_{x \in \mathcal{X}^M} f(x,y).
    \end{equation}
\end{lem}

Under Assumption \ref{asp:nonempty}, the scenario minimax problem \eqref{eq:scenario_minimax_def} can be framed as a \emph{simultaneous game}, in which the order of play is immaterial. In this convex--concave setting, the NE serves as a natural notion of stationarity.

\subsection{Robustness of an NE}   \label{subsec:convex}
Once an NE of the scenario minimax problem \eqref{eq:scenario_minimax_def} is found, a natural concern is its robustness against unseen samples outside the chosen multi-sample $\boldsymbol{\theta}^{M}$. We begin by reviewing the scenario approach, as adopted from \citet{campi2018wait}. The \emph{probability of violation} serves as a measure of robustness.
 
\begin{defn}   \label{def:probobility_of_violation}
    The probability of violation $V:\mathbb{R}^{p+q} \to [0,1]$ of a point $(x,y)$ is defined as 
    \begin{equation}  \nonumber
        V(x,y) : = \mathbb{P}\{ \theta \in \Theta : (x,y) \notin \mathcal{X}_{\theta} \times  \mathcal{Y}_{\theta} \}  .
    \end{equation}
\end{defn}

Definition \ref{def:probobility_of_violation} quantifies the risk that a solution $(x,y)$ to the scenario minimax problem \eqref{eq:scenario_minimax_def} is infeasible w.r.t. a new constraint $\mathcal{X}_{\Tilde{\delta}}$. Central concepts in the scenario approach also include \emph{support list} and \emph{complexity}, defined as follows.

\begin{defn}
    Given a multi-sample $\boldsymbol{\theta}^M$, a support list is a sequence of elements $\theta_{i_1},\dots,\theta_{i_k}$ of $\boldsymbol{\theta}^M$ with $i_1<\dots<i_k$ and $k \le M$ such that 
    \begin{enumerate}
        \item removing all elements of $\boldsymbol{\theta}^M$ except those in the sequence will not change the equilibrium;
        \item no more elements can be removed from the sequence while leaving the equilibrium unchanged.
    \end{enumerate}
\end{defn}

\begin{defn}
    The complexity of a multi-sample is the minimal cardinality among all its support lists.
\end{defn}

We make the following non-degeneracy assumption. This is considered a mild assumption in convex settings, a point discussed comprehensively by~\citet[Sec. 8]{campi2018wait} and~\citet[Sec. 1.3]{garatti2025non}.

\begin{assum}   [Non-degeneracy] \label{asp:non_dege}
    For any multi-sample $\boldsymbol{\theta}^{M}$ with $M \in \mathbb{N}$, with probability 1 over the random draws $\theta_1,\dots,\theta_M$, there exists a unique support list.
\end{assum}

Given Lemma \ref{lemma:convex_existence} and a tie-break rule ensuring uniqueness, e.g., by selecting the NE with minimal $\ell_2$-norm, a direct application of \citet[Th. 3]{campi2018wait} yields a robustness guarantee for the NE.

\begin{prop}   \label{prop:cc_robustness}
    Let Assumptions \ref{asp:nonempty} and \ref{asp:non_dege} hold, assume $f$ is convex--concave, and let $h(m)$ be any $[0,1]$-valued function for $m = 0,1,\dots,p+q$. Given a multi-sample $\boldsymbol{\theta}^{M}$, the NE $(x^*,y^*)$ to the scenario minimax problem \eqref{eq:scenario_minimax_def} satisfies
    \[
        \mathbb{P}^M \{ V(x^*,y^*) > h(S_M^{*}) \} \le \gamma^{*},
    \]
    where $S_M^{*}$ is the complexity of the multi-sample $\boldsymbol{\theta}^M$, and 
    \begin{equation}   \label{eq:gamma}
        \gamma^{*} = \inf_{\xi(\cdot) \in \boldsymbol{\mathrm{P}}_M} \xi(1)
    \end{equation}
    subject to
    \[
    \frac{1}{k!}\frac{\mathrm{d}^k}{\mathrm{d}t^k} \xi(t)  \ge \begin{pmatrix}
        M \\ k
    \end{pmatrix} t^{M-k} \cdot \mathds{1}\{t \in [0,1-h(k) ) \}
    \]
    for all $t \in [0,1]$ and all $k = 0,1,\dots,M$, where $\mathds{1} \{ \cdot \}$ denotes the indicator function, which equals 1 if the condition inside is true, and 0 otherwise; $\boldsymbol{\mathrm{P}}_M$ denotes the class of polynomials of degree $M$.
\end{prop}

After revisiting the convex--concave case, we now turn to the central contribution of this paper: establishing robustness results for nonconvex--nonconcave minimax problems. Notably, non-degeneracy in Assumption \ref{asp:non_dege} is generally difficult to verify, even in convex optimization, and this challenge is amplified in the context of games~\citep{fele2020probably}. Furthermore, the existence of an NE is no longer guaranteed once convexity is violated. 

Accordingly, we state our problem as follows.
\begin{prob}   \label{problem}
    For general nonconvex--nonconcave minimax problems, characterize the robustness of key equilibria.
\end{prob}

\section{Robustness of Nonconvex--Nonconcave Minimax Problems}    \label{sec:nonconvex}

In the nonconvex--nonconcave setting, demonstrating a robustness guarantee like Proposition \ref{prop:cc_robustness} is challenging because Assumption \ref{asp:non_dege}, a crucial non-degeneracy condition, typically fails to hold. This raises a key question: how can we guarantee the robustness of an equilibrium without the assumption of non-degeneracy? We find inspiration from \citet{garatti2025non}, who successfully removed the assumption of convexity in the scenario approach. Their main result can be stated as follows:

\begin{prop}     \label{thm:MP25}
    Given a multi-sample $\boldsymbol{\theta}^M$, if there is a map $\mathcal{M}_M : \Theta^{M} \to \mathcal{Z}_{\theta_M}$ satisfying the \emph{consistency property} defined as:
    \begin{enumerate}
        \item if $\theta_{i_1},\dots,\theta_{i_M}$ is a permutation of $\theta_{1},\dots,\theta_{M}$, then $\mathcal{M}_M (\theta_{i_1},\dots,\theta_{i_M}) = \mathcal{M}_M (\theta_{1},\dots,\theta_{M});$
        \item if $z_M^* \in \mathcal{Z}_{\theta_{M+i}}$ for all $i = 1,\dots,N$, then
        $
                 z_M^* = \mathcal{M}_M (\theta_{1},\dots,\theta_{M}) 
            = \mathcal{M}_{M+N} (\theta_{1},\dots,\theta_{{M+N}}) = z_{M+N}^* ;
        $
        \item if $z_M^* \notin \mathcal{Z}_{\theta_{M+i}}$ for at least one $i \in \{ 1, \dots, N\}$, then 
        $
                 z_M^* = \mathcal{M}_M (\theta_{1},\dots,\theta_{M})  
            \neq \mathcal{M}_{M+N} (\theta_{1},\dots,\theta_{{M+N}}) = z_{M+N}^*;
        $
    \end{enumerate}
    then we have
    \[
        \mathbb{P}^M \{ V(z_M^*) > h(S_M^{*}) \} \le \gamma^{*},
    \]
    where $h(k),k = 0,1,\dots,M$ is any $[0,1]$-valued function, and $\gamma^{*}$ is defined in \eqref{eq:gamma}.
\end{prop}

In the context of minimax problems, the map $\mathcal{M}_M$ serves as an algorithm to seek an equilibrium $z_M^* \in \mathcal{Z}_{\theta_M}$ based on the multi-sample $\boldsymbol{\theta}^M$. Properly defining stationarity for nonconvex--nonconcave minimax problems remains a fundamental challenge~\citep{li2025nonsmooth}. To navigate this difficulty, the remainder of this section investigates two distinct equilibrium concepts. Specifically, in Section~\ref{sec:epsilon}, we leverage the foundational result of Proposition~\ref{thm:MP25} to establish a probabilistic robustness guarantee for an $\varepsilon$-stationary point under convex strategy sets. Subsequently, in Section~\ref{sec:globalminimax}, we demonstrate that Proposition~\ref{thm:MP25} is fundamentally inapplicable to fully nonconvex settings, prompting us to derive an alternative robustness bound for a global minimax point.

\subsection{Robustness of an $\varepsilon$-Stationary Point}   \label{sec:epsilon}

We first examine the robustness of solutions to \eqref{eq:scenario_minimax_def} under Assumption~\ref{asp:nonempty}, namely, minimax problems with nonconvex--nonconcave payoffs and convex strategy sets. In this setting, a comprehensive description of various stationarity concepts can be found in~\citet{zhang2022optimality}. One approach is \emph{$\varepsilon$-stationary point}, which provides necessary conditions for a local NE; this concept is often analyzed under the Polyak-{\L}ojasiewicz (P{\L}) condition~\citep{yang2022faster} or the more general Kurdyka-{\L}ojasiewicz~(K{\L}) condition~\citep{zheng2023universal, li2025nonsmooth}. 

Following \cite{zheng2023universal}, we first define the concept of $\varepsilon$-stationary point.

\begin{defn}
A collective strategy $(\hat{x},\hat{y}) \in \mathcal{X}^M \times \mathcal{Y}^M$ is an $\varepsilon$-stationary point of \eqref{eq:scenario_minimax_def}, if 
    \begin{equation}\nonumber
    \begin{aligned}
    \operatorname{dist}\!\left( \boldsymbol{0}_p, \nabla_x f(\hat{x},\hat{y})
      + \mathcal{N}_{\mathcal{X}^M} (\hat{x}) \right) &\le \varepsilon, \\
    \operatorname{dist}\!\left( \boldsymbol{0}_q, -\nabla_y f(\hat{x},\hat{y})
      + \mathcal{N}_{\mathcal{Y}^M} (\hat{y}) \right) &\le \varepsilon,
    \end{aligned}
    \end{equation}
where $\mathcal{N}_{\mathcal{S}}$ is the normal cone operator associated with a set $\mathcal{S} \subseteq \mathbb{R}^d$, and $\operatorname{dist}(z,\mathcal{S}) := \inf_{v \in \mathcal{S}} \| z - v \|$ denotes the distance from $z \in \mathbb{R}^d$ to $\mathcal{S}$.
\end{defn}

Under Assumption \ref{asp:nonempty}, often in conjunction with the P{\L} or K{\L} condition, convergence to an $\varepsilon$-stationary point can often be achieved, typically via the gradient descent-ascent~(GDA) algorithm and its variants~\citep{li2025nonsmooth,yang2022faster}. Despite its importance, the robustness of the resulting $\varepsilon$-stationary point remains largely unexplored, and this work aims to fill this research gap.

Quantifying the robustness of $\varepsilon$-stationary points is challenging in the nonconvex--nonconcave case, where the complex geometry and potential for degeneracy make direct analysis intractable. Our core strategy is to first prove the existence of an $\varepsilon$-stationary point under Assumption \ref{asp:nonempty}, and then establish the consistency property introduced in Proposition \ref{thm:MP25} and in that way characterize the stability of $\varepsilon$-stationary points against sample changes. Note that the result holds even without a non-degeneracy assumption. 
\begin{thm}    \label{thm:epsilon}
    Let Assumption \ref{asp:nonempty} hold. For a given multi-sample $\boldsymbol{\theta}^M$, there exists an $\varepsilon$-stationary point $(\hat{x},\hat{y})$ of \eqref{eq:scenario_minimax_def} for some $\varepsilon > 0$ such that
    \[
        \mathbb{P}^M \{ V(\hat{x},\hat{y}) > h(S_M^{*}) \} \le \gamma^{*},
    \]
    where $\gamma^{*}$ is defined in \eqref{eq:gamma}.
\end{thm}
\begin{pf}
    The $\varepsilon$-stationary point's invariance to permutation of the multi-sample $\boldsymbol{\theta}^M$ means that condition (1) of the consistency property is naturally met. 
    
     To make the dependence on the sample set explicit for a fixed $\boldsymbol{\theta}^M$, define the stationary residual
    \begin{equation} \nonumber
    \begin{aligned}
        r^M(x,y) := \max \big\{
        &\operatorname{dist}\!\left( \boldsymbol{0}_p, \nabla_x f(x,y)
        + \mathcal{N}_{\mathcal{X}^M} (x) \right), \\
        &\operatorname{dist}\!\left( \boldsymbol{0}_q, -\nabla_y f(x,y)
        + \mathcal{N}_{\mathcal{Y}^M} (y) \right) \big\}.
    \end{aligned}
    \end{equation}
    The continuous differentiability of $f$ implies the continuity of both $\nabla_x f$ and $\nabla_y f$. By \citet[Prop. 6.6]{rockafellar1998variational}, relative to $\mathcal{X}^M$, the set-valued mapping $x \mapsto \mathcal{N}_{\mathcal{X}^M}(x)$ is outer semicontinuous with closed values.
    Therefore, \citet[Prop. 5.11]{rockafellar1998variational} guarantees that for fixed $y$, the map
    $
    x \mapsto \operatorname{dist}\!\left( \boldsymbol{0}_p, \nabla_x f(x,y)
    + \mathcal{N}_{\mathcal{X}^M} (x) \right)
    $
    is lower semicontinuous (lsc) relative to $\mathcal{X}^M$, and similarly for the $y$-component. Consequently, since $r^M(x,y)$ is the maximum of two lsc functions, it follows directly that $r^M$ is lsc relative to $\mathcal{X}^M \times \mathcal{Y}^M$. Under Assumption \ref{asp:nonempty}, \citet[Th. 1.9]{rockafellar1998variational} guarantees that there exists $(\hat{x}_M,\hat{y}_M) \in \arg \min_{(x,y) \in \mathcal{X}^M \times \mathcal{Y}^M} r^M(x,y)$ such that $r^M(\hat{x}_M,\hat{y}_M) = \varepsilon$.

    Now consider augmenting $\boldsymbol{\theta}^M$ by an additional multi-sample $\boldsymbol{\theta}^N$. Observe that both $\mathcal{X}^{M+N}$ and $\mathcal{X}^M $ are convex, and the subset relation $\mathcal{X}^{M+N} \subseteq \mathcal{X}^M $ leads to
    \begin{equation}  \nonumber
        \mathcal{N}_{\mathcal{X}^M} (x) 
        \subseteq
        \mathcal{N}_{\mathcal{X}^{M+N}} (x),\ 
        \forall x \in \mathcal{X}^M,
    \end{equation}
    and similarly for the $y$-component. Hence, for every $(x,y)$ in the common domain $\mathcal{X}^M \times \mathcal{Y}^M$,
    $
        \operatorname{dist}\big(\mathbf{0}_p,\nabla_x f(x,y)+\mathcal N_{\mathcal{X}^{M+N}}(x)\big) 
        \le  \operatorname{dist}\big(\mathbf{0}_p,\nabla_x f(x,y)+\mathcal N_{\mathcal{X}^M}(x)\big),
    $
    and an analogous inequality holds for the $y$-component. It follows that for all $(x,y)$,
    $
        r^{M+N}(x,y) \le r^{M}(x,y).
    $
 Take the pair
    \[
    (\hat x_{M+N},\hat y_{M+N})\in\arg\min_{(x,y) \in \mathcal{X}^{M
    +N} \times \mathcal{Y}^{M+N}} r^{M+N}(x,y)
    \] 
    as a minimizer for the augmented multi-sample $\boldsymbol{\theta}^{M+N}$. If for all $j\in\{1,\dots,N\}$ it holds that
    $
    (\hat x_M,\hat y_M) \in \mathcal{X}^{M+j} \times \mathcal{Y}^{M+j},
    $
    then in particular 
    $
    (\hat x_M,\hat y_M) \in \mathcal{X}^{M
    +N} \times \mathcal{Y}^{M+N}
    $ and therefore
    $$
        r^{M+N}(\hat x_{M+N},\hat y_{M+N}) 
        \!\le \!r^{M+N}(\hat x_M,\hat y_M)
        \!\le \!r^{M}(\hat x_M,\hat y_M) 
        \!=\! \varepsilon. 
    $$
    This shows that the $\varepsilon$-stationary point remains valid after adding new appropriate samples;
    i.e., condition (2) of the consistency property holds.

    Conversely, if for some $j\in\{1,\dots,N\}$ the point $(\hat x_M,\hat y_M)$ fails to belong to
    $\mathcal{X}^{M+j} \times\mathcal{Y}^{M+j}$, then in particular we have
    $(\hat x_M,\hat y_M)\notin\mathcal{X}^{M
    +N} \times \mathcal{Y}^{M+N}.$ Recall that 
    $(\hat x_{M+N},\hat y_{M+N})\in\mathcal{X}^{M
    +N} \times \mathcal{Y}^{M+N}.$ Thus, we conclude that
    $(\hat x_M,\hat y_M)\neq(\hat x_{M+N},\hat y_{M+N}),$
    thereby verifying condition (3) of the consistency property.
    
Combining the above observations with Proposition~\ref{thm:MP25} completes the proof. \hfill $\qed$
\end{pf}

Theorem \ref{thm:epsilon} establishes a generalization guarantee of an $\varepsilon$-stationary point to problem \eqref{eq:scenario_minimax_def}. Given a solution $(\hat x_M,\hat y_M)$ obtained via a suitable algorithm, such as the one proposed in \citet{zheng2023universal}, Theorem~\ref{thm:epsilon} provides a probabilistic upper bound on its generalization error. Notably, while our proof starts from the existence of an $\varepsilon$-stationary point, the resulting robustness bound is algorithm-independent and valid for any point meeting this condition. For instance, one could first determine a desired small $\varepsilon$ and employ a suitable algorithm to reach this threshold. Consequently, treating this achieved residual as $\varepsilon$, the remainder of the theoretical guarantees follows naturally.

From a practical perspective, one typically fixes a high confidence level $1 - \beta$ (i.e., very close to 1) and, in return, desires a deterministic upper bound $h(S_M^{*})$ for the probability of violation $V(\hat x_M,\hat y_M)$. Since the function $h(\cdot)$ as in Proposition~\ref{thm:MP25} and Theorem~\ref{thm:epsilon} is indefinite for a fixed $\gamma^*$, there may exist infinitely many solutions $h(k)$ to \eqref{eq:gamma}. A practical corollary resolves this by setting $\xi (t) = \frac{\beta}{M} \sum_{n=0}^{M-1} t^n$~\citep[Th. 4]{garatti2025non}, as follows.

\begin{cor}   \label{coro:unique_epsilon}
    Under Assumption \ref{asp:nonempty}, for a given $\beta \in (0,1)$ and a multi-sample $\boldsymbol{\theta}^{M}$, the $\varepsilon$-stationary point $(\hat{x},\hat{y})$ of \eqref{eq:scenario_minimax_def} satisfies
    \begin{equation}  \nonumber
        \mathbb{P}^M \{ V(\hat{x},\hat{y}) > g(S_M^{*}) \} \le \beta,
    \end{equation}
    where function $g(k)$ is defined as
    \begin{equation}  \label{eq:unqiue_bound}
        g(k) = 
        \begin{cases} 
            1 - t(k), \quad &\text{if } k = 0 ,1 , \dots, M-1, \\
            1, &\text{if } k = M,
        \end{cases}
    \end{equation}
    with $t(k)$ being the unique solution in the interval $(0,1)$ of 
    \begin{equation}  \nonumber
        \frac{\beta}{M} \sum_{m=k}^{M-1} \binom{m}{k} t^{m-k} - \binom{M}{k} t^{M-k} = 0.
    \end{equation}
\end{cor}

Equivalently, the result in Corollary~\ref{coro:unique_epsilon} can be stated in terms of the complementary probability:
\begin{equation}  \nonumber
        \mathbb{P}^M \{ V(\hat{x},\hat{y}) \le g(S_M^{*}) \} \ge 1 - \beta.
\end{equation}
This probabilistic bound gives a powerful robustness guarantee for an obtained $\varepsilon$-stationary point $(\hat{x},\hat{y})$ from a multi-sample $\boldsymbol{\theta}^M$. By first selecting an arbitrarily high confidence level $1 - \beta$ and then computing the complexity $S_M^{*}$ of $\boldsymbol{\theta}^M$, one can uniquely determine the upper bound $ g(S_M^{*}) $ using \eqref{eq:unqiue_bound}. That is, we have at least $1 - \beta$ confidence to ensure that $ V(\hat{x},\hat{y}) $ will not exceed this computed and unique $ g(S_M^{*}) $. Note that the probability of violation $ V(\hat{x},\hat{y}) $ serves as a measure of robustness: the smaller its value, the more robust the $\varepsilon$-stationary point $(\hat{x},\hat{y})$ is. Therefore, Corollary~\ref{coro:unique_epsilon} formalizes this probabilistic robustness guarantee.

\subsection{Robustness of a Global Minimax Point}       \label{sec:globalminimax}

Let us consider the minimax problem \eqref{eq:scenario_minimax_def} with nonconvex--nonconcave payoffs and nonconvex strategy sets. In Section \ref{sec:epsilon}, we investigate the robustness of an $\varepsilon$-stationary point by playing a simultaneous game. However, in this simultaneous setting, it is not straightforward to extend the result of Theorem~\ref{thm:epsilon} to minimax problems with nonconvex strategy sets. This is because the monotonicity of the stationary residual w.r.t. the number of scenarios is difficult to maintain when the convexity assumption on the strategy sets (Assumption~\ref{asp:nonempty}) is violated.

Interestingly, the GANs framework implements a sequential two-player game: it first optimizes the discriminator $D$ while keeping the generator $G$ fixed, and then updates $G$ using the optimized $D$. More generally, adversarial training describes the minimax problem \eqref{eq:scenario_minimax_def} in the frame of a \emph{sequential game}. These cases lead to the equilibrium concept known as \emph{global minimax point}, which captures global optimality and is widely employed in machine learning~\citep{jin2020local,zhang2022optimality,chen2024approaching}. Intuitively, the min-player acts as the leader, first determining their optimal strategy, while the max-player then acts as the follower, choosing a best response to the min-player's choice. For a game-theoretic perspective, this setup corresponds to a Stackelberg game.

Following \cite{jin2020local}, we begin by defining the concept of global minimax point. In this paper, we focus exclusively on a global minimax point of \eqref{eq:scenario_minimax_def}, since any global maximin point corresponds to its global minimax point with the payoff $-f(y,x)$.

\begin{defn}  \label{def:minimax}
    A collective strategy $(x^{\star} , y^{\star}) \in \mathcal{X}^M \times \mathcal{Y}^M$ is a global minimax point of \eqref{eq:scenario_minimax_def}, if for all $(x , y)$:
    \begin{equation}   \nonumber
        f(x^{\star},y) \le f(x^{\star} , y^{\star}) \le \max_{y' \in \mathcal{Y}^M} f(x,y').
    \end{equation}
\end{defn}

The next assumption removes all convexity requirements, including the strategy sets originally assumed to be convex in Assumption \ref{asp:nonempty}.

\begin{assum}  [Strategy nonconvexity]   \label{asp:non-convex}
    $\mathcal{X}_{\theta}$ and $\mathcal{Y}_{\theta}$ are compact and nonempty for all $\theta \in \Theta$.
\end{assum}

\begin{lem} [{\citealp{jin2020local}}]   \label{lemma:global_existence}
    Under Assumption \ref{asp:non-convex}, there exists at least one global minimax point $(x^{\star} , y^{\star})$ of the scenario minimax problem~\eqref{eq:scenario_minimax_def}.
\end{lem}

\begin{rem}
    It is important to note that in the absence of further structural assumptions beyond Assumption~\ref{asp:non-convex}, the consistency property for a global minimax point does not hold in general (Proposition~\ref{thm:MP25}).
\end{rem}


To see that, let us define the envelope functions $$\phi_M(x) := \max_{y' \in \mathcal{Y}^M} f(x,y'),\ \phi_{M+N} (x) := \max_{y' \in \mathcal{Y}^{M+N}} f(x,y').$$ By Definition~\ref{def:minimax}, given a multi-sample $\boldsymbol{\theta}^M$, the global minimax point $(x_M^\star,y_M^\star)$ satisfies $\phi_M (x_M^\star) \le \phi_M(x),\ \forall x \in \mathcal{X}^M$. Augmenting $\boldsymbol{\theta}^M$ by an additional multi-sample $\boldsymbol{\theta}^N$, and assuming $(x_M^\star,y_M^\star) \in \mathcal{X}^{M+N} \times \mathcal{Y}^{M+N}$, we have $\phi_{M+N} (x_M^\star) = \max_{y' \in \mathcal{Y}^{M+N}} f(x_M^\star,y') = \phi_M (x_M^\star)$, where the last equality holds because $y_M^\star$ is the maximizer over the larger feasible set $\mathcal{Y}^M$, and it remains feasible in the smaller set $\mathcal{Y}^{M+N}$.

The consistency property holds if the original $x_M^\star$ satisfies $\phi_{M+N} (x_M^\star) \le \phi_{M+N} (x),\ \forall x \in \mathcal{X}^{M+N}. $ However, the two available conditions: 1) $\phi_{M+N} (x_M^\star) = \phi_M (x_M^\star) \le \phi_M(x),\ \forall x \in \mathcal{X}^{M+N}$; 2) $\phi_{M+N} (x) \le \phi_M (x),\ \forall x \in \mathcal{X}^{M+N}$, cannot conclude $\phi_{M+N} (x_M^\star) \le \phi_{M+N} (x)$ in general. Namely, the global minimax point is not necessarily preserved after adding appropriate samples. Hence, the consistency property does not hold. As a relaxation, we establish its probabilistic robustness by leveraging an earlier foundational result~\citep{campi2018general}.

\begin{thm}   \label{thm:nonconvex}
    Let Assumption \ref{asp:non-convex} hold. For a given multi-sample $\boldsymbol{\theta}^M$, the global minimax point $(x^\star,y^\star)$ of the scenario minimax problem \eqref{eq:scenario_minimax_def} satisfies
    \begin{equation}   \nonumber
        \mathbb{P}^M \{ V(x^\star,y^\star) > \epsilon(S_M^{*}) \} \le \beta,
    \end{equation}
    where $\epsilon (k) = 1 - \exp \left(  \frac{1}{M-k} \ln \left( \frac{\beta}{ M \binom{M}{k} } \right) \right) $ for $k = 0, 1, \dots, M-1$, with $\epsilon (M) = 1$.
\end{thm}
\begin{pf}
    Lemma~\ref{lemma:global_existence}, proven by the extreme value theorem, ensures the existence of a global minimax point, but not its uniqueness. To overcome this issue, we equip a suitable tie-breaking rule to $\mathcal{M}_M$, such as choosing a point $(x^\star,y^\star)$ with minimal $\ell_2$-norm. Thus, $\mathcal{M}_M (\theta_{1},\dots,\theta_{M}) = (x^\star,y^\star) \in \mathcal{X}^M \times \mathcal{Y}^M =  \bigcap_{i = 1}^{M} \mathcal{X}_{\theta_{i}} \times  \bigcap_{i = 1}^{M} \mathcal{Y}_{\theta_{i}}$, which implies $\mathcal{M}_M (\theta_{1},\dots,\theta_{M}) \in \mathcal{X}_{\theta_i} \times \mathcal{Y}_{\theta_i}$ for all $i \in \{ 1,\dots,M \}$. Applying \citet[Th. 1]{campi2018general} yields $\sum_{k=0}^{M-1} \binom{M}{k} (1 - \epsilon(k) )^{M-k} = \beta$, with $\epsilon(M) = 1$. As indicated in~\cite{campi2018general}, setting each term in the summation equal to $\beta / M$ leads to the exponential-logarithmic form of $\epsilon(k)$ by taking the logarithm on both sides. This completes the proof. \hfill $\qed$
\end{pf}

While Theorem \ref{thm:epsilon} focuses on a local $\varepsilon$-stationary point, Theorem \ref{thm:nonconvex} shows that a different high-probability generalization bound also holds for a global minimax point, even in fully nonconvex settings. By calculating $\epsilon(S_M^{*})$--which is typically small in practice--one can assert, with probability at least $1 - \beta$, that the probability of violation $V(x^\star,y^\star)$ does not exceed $\epsilon(S_M^{*})$. However, as a natural trade-off for achieving global optimality in fully nonconvex problems, the generalization bound derived in Theorem~\ref{thm:nonconvex} is inherently more conservative than that in Theorem~\ref{thm:epsilon}.

\section{Numerical Experiment}   \label{sec:Numerical Example}

In this section, we validate the theoretical results by considering a nonconvex--nonconcave minimax optimization problem with bilinear interaction~\citep{grimmer2023landscape} under uncertainty
\begin{equation}  \nonumber
\begin{aligned}
    \min_{x \in \mathcal{X}_{\theta}} \max_{y \in \mathcal{Y}_{\theta}}f(x,y) 
    & = (x + 1)(x + 3)(x - 1)(x - 3) \\
    & \quad + 10 xy \\
    & \quad  - (y + 1)(y + 3)(y - 1)(y - 3),
\end{aligned}
\end{equation}
where the uncertainty sets $\mathcal{X}_{\theta} = \mathcal{Y}_{\theta} = [-2\theta, -\theta] \cup [\theta, 2\theta]$ with $\theta$ uniformly distributed in $\Theta = [1.0, 1.5]$.

Given the scenario minimax problem associated with a multi-sample $\boldsymbol{\theta}^{M}$, we compute $\phi(x) = \max_y f(x,y)$ and select the min-player's strategy as
$
    x^\star_M = \arg \min_x \phi(x) = \arg \min_x \max_y f(x,y),
$
while the max-player chooses
$
    y^\star_M = \arg \max_y f(x^\star_M,y).
$
With this minimax point $(x^\star_M,y^\star_M)$, we aim to ensure that
$
    \mathbb{P}\{ \theta \in \Theta : (x^\star_M,y^\star_M) \notin \mathcal{X}_{\theta} \times \mathcal{Y}_{\theta} \} > \epsilon (S_M^{*})
$
holds with probability at most $\beta$, where the complexity $S_M^{*}$ is evaluated using the greedy algorithm described in \citet[Sec. 2]{campi2018general}.

To empirically validate this bound $\epsilon (S_M^{*})$, a Monte Carlo simulation with $R = 200$ repetitions is performed for different number of scenarios, $M$, ranging from $1$ to $100$, and for $\beta = 0.01$. In each repetition, a set of $M$ scenarios is sampled, the global minimax point $(x^\star_M,y^\star_M)$ is computed, and its complexity $S_M^{*}$ is recorded to determine the theoretical bound $\epsilon (S_M^{*})$. Following this, the solution is subject to an out-of-sample test against a large independent set of $N_{\text{test}} = 10^4$ test scenarios. This test yields the empirical probability of violation
$
    \hat{V}_M(x^\star_M,y^\star_M) = \frac{1}{N_{\text{test}}} \sum_{i=1}^{N_{\text{test}}} \mathds{1}\{ (x^\star_M,y^\star_M) \notin \mathcal{X}_{\theta_i} \times \mathcal{Y}_{\theta_i} \},
$
for that repetition. The average empirical probability of violation over $R$ repetitions in the Monte Carlo simulation is denoted by $\Bar{V}_M(x^\star_M,y^\star_M)$.

\begin{figure}[t]
\centering
\includegraphics[width=8.8cm]{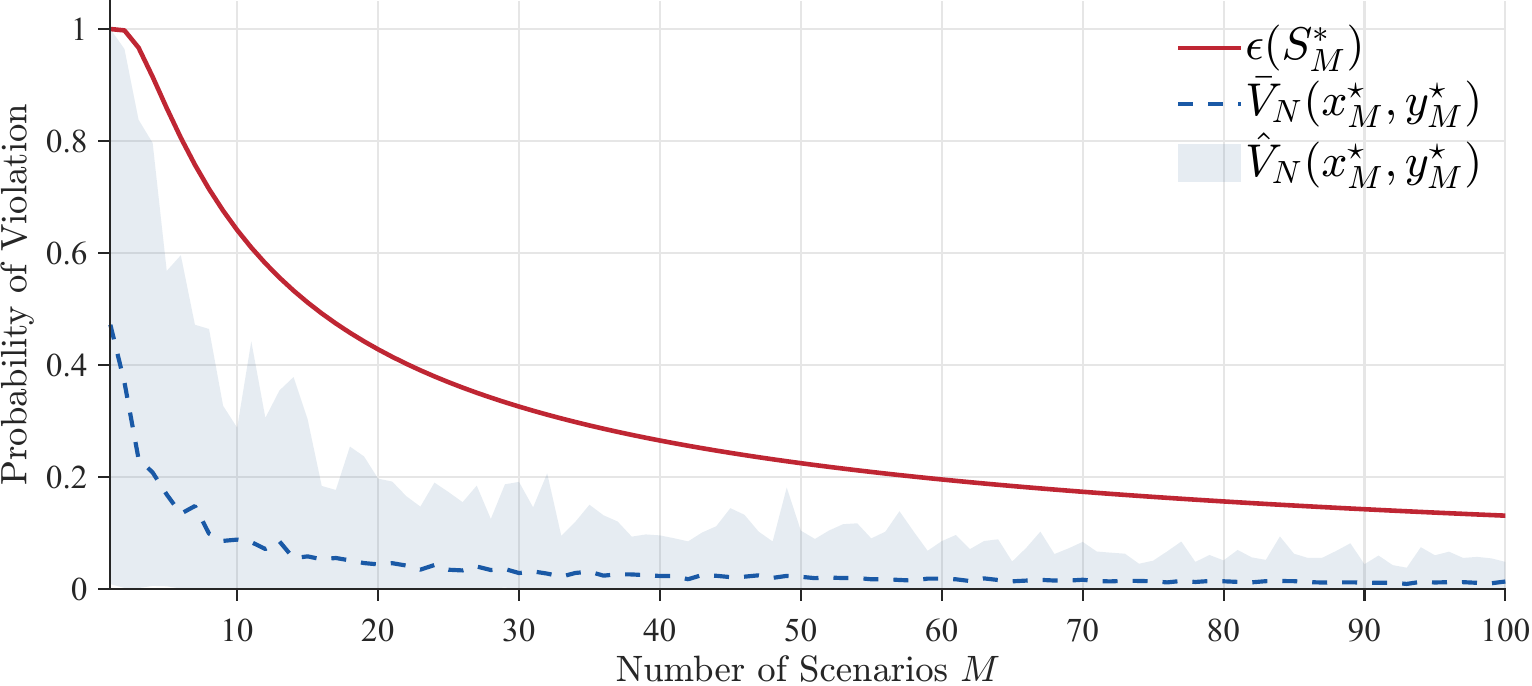}
\caption{The probability of violation depends on the number of scenarios for $\beta = 0.01$. The figure compares the theoretical bound $\epsilon (S_M^{*})$ (red solid line) with the range of all possible empirical probabilities of violation $\hat{V}_M(x^\star_M, y^\star_M)$ (light purple shaded area) and the average empirical probability of violation $\bar{V}_M(x^\star_M, y^\star_M)$ (blue solid line) across different numbers of scenarios.}
\label{fig:the_prob_of_vio_0.01}
\end{figure}

Unlike the probability of violation, the complexity $S_M^{*}$ is identical across all Monte Carlo repetitions for any $M$. This is due to the special structure of this numerical example, specifically the bilinear payoff and interval uncertainty. Therefore, $\epsilon (S_M^{*})$ in Fig.~\ref{fig:the_prob_of_vio_0.01} remains constant for a given $M$. Moreover, we observe that the empirical probability of violation can be exactly zero, implying the absence of a strictly positive lower bound. By fixing $\beta = 0.01$ (a $99\%$ confidence level), we expect the empirical violation $\hat{V}_M(x^\star_M,y^\star_M)$ to exceed the theoretical bound $\epsilon (S_M^{*})$ in no more than $1\%$ of the repetitions. In Fig.~\ref{fig:the_prob_of_vio_0.01}, our numerical results are consistent with this guarantee, as the bound was never violated. This provides a numerical verification of Theorem~\ref{thm:nonconvex}.

\section{CONCLUSIONS}  \label{sec:conclusion}
This paper developed a data-driven and distribution-free framework to establish probabilistic robustness guarantees for an $\varepsilon$-stationary point and a global minimax point in general nonconvex--nonconcave minimax problems. In particular, we obtained a tight robustness guarantee for a local equilibrium through an $\varepsilon$-stationary point under convex strategy sets. Conversely, under fully nonconvex strategy sets, we demonstrated that global minimax points adhere to an alternative, relaxed robustness bound. A critical direction for future research was to theoretically establish a tight probabilistic bound for global minimax points, a possibility strongly hinted at by our numerical validation of the tight bound in Proposition~\ref{thm:MP25}. We also identified the extension of this framework to nonconvex multiplayer games as the most challenging direction for future research.

\bibliography{ifacconf}             

@article{li2025nonsmooth,
  title={Nonsmooth nonconvex--nonconcave minimax optimization: Primal--dual balancing and iteration complexity analysis},
  author={Li, Jiajin and Zhu, Linglingzhi and So, Anthony Man-Cho},
  journal={Mathematical Programming},
  pages={1--51},
  year={2025},
  publisher={Springer}
}

@article{zhang2022optimality,
  title={Optimality and stability in non-convex smooth games},
  author={Zhang, Guojun and Poupart, Pascal and Yu, Yaoliang},
  journal={Journal of Machine Learning Research},
  volume={23},
  number={35},
  pages={1--71},
  year={2022}
}

@inproceedings{jin2020local,
  title={What is local optimality in nonconvex-nonconcave minimax optimization?},
  author={Jin, Chi and Netrapalli, Praneeth and Jordan, Michael},
  booktitle={International Conference on Machine Learning},
  pages={4880--4889},
  year={2020},
}

@article{garatti2025non,
  title={Non-convex scenario optimization},
  author={Garatti, Simone and Campi, Marco C},
  journal={Mathematical Programming},
  volume={209},
  number={1},
  pages={557--608},
  year={2025},
  publisher={Springer}
}

@article{sion1958general,
  title={On general minimax theorems},
  author={Sion, Maurice},
  journal={Pacific Journal of Mathematics},
  volume={8},
  number={1},
  pages={171--176},
  year={1958},
  publisher={Mathematical Sciences Publishers}
}

@article{campi2018wait,
  title={Wait-and-judge scenario optimization},
  author={Campi, Marco C and Garatti, Simone},
  journal={Mathematical Programming},
  volume={167},
  number={1},
  pages={155--189},
  year={2018},
  publisher={Springer}
}

@article{fele2020probably,
  title={Probably approximately correct {N}ash equilibrium learning},
  author={Fele, Filiberto and Margellos, Kostas},
  journal={IEEE Transactions on Automatic Control},
  volume={66},
  number={9},
  pages={4238--4245},
  year={2020},
  publisher={IEEE}
}

@book{bacsar2008h,
  title={H-infinity Optimal Control and Related Minimax Design Problems: A Dynamic Game Approach},
  author={Ba{\c{s}}ar, Tamer and Bernhard, Pierre},
  edition  = {2nd},
  year={2008},
  publisher={Springer Science \& Business Media}
}

@book{Sutton2018,
  author = {Sutton, Richard S. and Barto, Andrew G.},
  edition = {2nd},
  publisher = {MIT Press},
  title = {Reinforcement Learning: An Introduction},
  year = {2018}
}

@inproceedings{goodfellow2014generative,
  author = {Goodfellow, Ian and Pouget-Abadie, Jean and Mirza, Mehdi and Xu, Bing and Warde-Farley, David and Ozair, Sherjil and Courville, Aaron and Bengio, Yoshua},
  booktitle = {Advances in Neural Information Processing Systems},
  pages = {2672--2680},
  title = {Generative adversarial nets},
  year = 2014
}

@article{campi2018general,
  title={A general scenario theory for nonconvex optimization and decision making},
  author={Campi, Marco Claudio and Garatti, Simone and Ramponi, Federico Alessandro},
  journal={IEEE Transactions on Automatic Control},
  volume={63},
  number={12},
  pages={4067--4078},
  year={2018},
  publisher={IEEE}
}

@article{fan1953minimax,
  title={Minimax theorems},
  author={Fan, Ky},
  journal={Proceedings of the National Academy of Sciences},
  volume={39},
  number={1},
  pages={42--47},
  year={1953}
}

@article{v1928theorie,
  title={Zur {T}heorie der {G}esellschaftsspiele},
  author={von Neumann, J.},
  journal={Mathematische Annalen},
  volume={100},
  number={1},
  pages={295--320},
  year={1928},
  publisher={Springer}
}

@inproceedings{madry2018towards,
  title={Towards Deep Learning Models Resistant to Adversarial Attacks},
  author={Madry, Aleksander and Makelov, Aleksandar and Schmidt, Ludwig and Tsipras, Dimitris and Vladu, Adrian},
  booktitle={International Conference on Learning Representations},
  pages={4138--4160},
  year={2018}
}

@article{assif2020scenario,
  title={Scenario approach for minmax optimization with emphasis on the nonconvex case: Positive results and caveats},
  author={Assif, Mishal and Chatterjee, Debasish and Banavar, Ravi},
  journal={SIAM Journal on Optimization},
  volume={30},
  number={2},
  pages={1119--1143},
  year={2020},
  publisher={SIAM}
}

@article{doi:10.1137/130928546,
author = {Car\`{e}, A. and Garatti, S. and Campi, M. C.},
title = {Scenario Min-Max Optimization and the Risk of Empirical Costs},
journal = {SIAM Journal on Optimization},
volume = {25},
number = {4},
pages = {2061-2080},
year = {2015},
}

@article{chen2021distributed,
  title={Distributed algorithm for $\varepsilon$-generalized {N}ash equilibria with uncertain coupled constraints},
  author={Chen, Guanpu and Ming, Yang and Hong, Yiguang and Yi, Peng},
  journal={Automatica},
  volume={123},
  pages={109313},
  year={2021},
  publisher={Elsevier}
}

@inproceedings{zheng2023universal,
  title={Universal gradient descent ascent method for nonconvex-nonconcave minimax optimization},
  author={Zheng, Taoli and Zhu, Linglingzhi and So, Anthony Man-Cho and Blanchet, Jos{\'e} and Li, Jiajin},
  booktitle={Advances in Neural Information Processing Systems},
  pages={54075--54110},
  year={2023}
}

@inproceedings{yang2022faster,
  title={Faster single-loop algorithms for minimax optimization without strong concavity},
  author={Yang, Junchi and Orvieto, Antonio and Lucchi, Aurelien and He, Niao},
  booktitle={International Conference on Artificial Intelligence and Statistics},
  pages={5485--5517},
  year={2022}
}

@book{rockafellar1998variational,
  title={Variational Analysis},
  author={Rockafellar, R Tyrrell and Wets, Roger JB},
  year={1998},
  publisher={Springer}
}

@article{chen2024approaching,
  title={Approaching the global {N}ash equilibrium of non-convex multi-player games},
  author={Chen, Guanpu and Xu, Gehui and He, Fengxiang and Hong, Yiguang and Rutkowski, Leszek and Tao, Dacheng},
  journal={IEEE Transactions on Pattern Analysis and Machine Intelligence},
  volume={46},
  number={12},
  year={2024},
  pages={10797–10813},
  publisher={IEEE}
}

@inproceedings{fochesato2023generalized,
  title={Generalized uncertain {N}ash games: Reformulation and robust equilibrium seeking},
  author={Fochesato, Marta and Fabiani, Filippo and Lygeros, John},
  booktitle={Proceedings of the 21st European Control Conference (ECC)},
  pages={1--6},
  year={2023},
  organization={IEEE}
}

@article{xu2023algorithm,
  title={Algorithm design and approximation analysis on distributed robust game},
  author={Xu, Gehui and Chen, Guanpu and Qi, Hongsheng},
  journal={Journal of Systems Science and Complexity},
  volume={36},
  number={2},
  pages={480--499},
  year={2023},
  publisher={Springer}
}

@article{grimmer2023landscape,
  title={The landscape of the proximal point method for nonconvex--nonconcave minimax optimization},
  author={Grimmer, Benjamin and Lu, Haihao and Worah, Pratik and Mirrokni, Vahab},
  journal={Mathematical Programming},
  volume={201},
  number={1},
  pages={373--407},
  year={2023},
  publisher={Springer}
}

@article{chen2025inverse,
  title={Inverse learning of black-box aggregator for robust {N}ash equilibrium},
  author={Chen, Guanpu and Xu, Gehui and He, Fengxiang and Tao, Dacheng and Parisini, Thomas and Johansson, Karl Henrik},
  journal={IEEE Transactions on Automatic Control},
  volume={71},
  number={5},
  pages={3357--3364},
  year={2026},
  publisher={IEEE}
}
                                                   







\end{document}